\documentclass[prl,twocolumn,showpacs]{revtex4}
\usepackage{amsmath}
\newcommand{\ep}{\varepsilon}
\newcommand{\rmd}{\text{d}}
\newcommand{\dfracd}[2]{\dfrac{\rmd #1}{\rmd #2}}
\newcommand{\dfracp}[2]{\dfrac{\partial #1}{\partial #2}}
\newcommand{\lessless}{\hspace*{-0.3em}<\hspace*{-0.4em}<\hspace*{-0.3em}}
\newcommand{\moremore}{\hspace*{-0.3em}>\hspace*{-0.4em}>\hspace*{-0.3em}}

\begin{document}
\title{New renormalization procedure for perturbed ordinally
  differential equations with given initial condition and prescribed
  secular terms}
\author{YAMAGUCHI Y. Yoshiyuki}
\email{yyama@amp.i.kyoto-u.ac.jp}
\homepage{http://yang.amp.i.kyoto-u.ac.jp/~yyama/}
\affiliation{Department of Applied Mathematics and Physics, 
  Kyoto University, Kyoto, 606-8501, Japan}
\date{\today}
\begin{abstract}
  For perturbed ordinally differential equations,
  a procedure of renormalization group method is proposed.
  To uniquely obtain renormalized solutions for given initial
  conditions, the procedure assumes that
  the extra integral constants yielded by perturbative expansion
  are determined as they cancel out all the prescribed secular terms.
  A sufficient condition for the renormalized solutions
  coinciding with exact solutions is also stated.
\end{abstract}
\maketitle

To obtain solutions to perturbed differential or discrete equations,
a perturbative expansion is usually applied.
But naive perturbation often yields secular terms due to resonance
phenomena,
and the secular terms prevent us from getting approximate but global
solutions.
The renormalization group method \cite{chen-94} is one of singular
perturbation techniques,
and gives the approximate but global solutions to a wide variety of 
differential \cite{chen-96,goto-99} and discrete
\cite{kunihiro-97,goto-01a} equations.
The method removes the secular terms
by renormalizing integration constants contained in a solution to an
unperturbed equation.
Temporal evolution of the renormalized constants is governed by a 
renormalization group equation,
which is mathematically regarded as an envelope equation for a family
of naive solutions \cite{kunihiro-95}.

However, a perturbative solution up to the $k$-th order 
of a small parameter $\ep$ contains $(k+1)$ times as many 
integration constants as the exact solution does
because of the presence of undetermined constants
appearing from the homogeneous solution from each order,
and hence we cannot obtain the unique renormalized solution
from a given initial condition without any assumptions.
In this letter, we introduce an assumption,
and propose a systematic procedure to uniquely obtain the renormalized
solutions both for oscillating naive solutions and for non-oscillating 
ones.
We study perturbation up to the first order of $\ep$ 
throughout this letter. 

Consider the perturbed differential equation of the form
\begin{equation}
  \label{eq:original-eq}
  \dfracd{x}{t} = f(x) + \ep g(x), 
  \quad x \in R^{n},
\end{equation}
where $x$ is a $n$-dimensional column vector,
$\ep$ is the small parameter, i.e. $|\ep| \lessless 1$,
and the unperturbed part of this equation, $\rmd x/\rmd t=f(x)$,
is integrable.
To obtain a naive solution,
$x$ is expanded as a positive power series of $\ep$;
$x=x^{(0)}+\ep x^{(1)}+\cdots$.
The equation of the zeroth order, the unperturbed equation, is
\begin{displaymath}
  \dfracd{x^{(0)}}{t} = f(x^{(0)}),
\end{displaymath}
and the general solution to this equation is denoted by
$x^{(0)}=\phi^{(0)}(t;A),\ A\in R^{n}$,
where each element of the $n$-dimensional column vector $A$ is
an integration constant. 
We assume the function $\phi^{(0)}$ is smooth with respect to both 
$t$ and $A$.
Denoting the Jacobian of $f$ by $Df$,
the first order equation is expressed as
\begin{equation}
  \label{eq:1st-eq}
  \dfracd{x^{(1)}}{t} = Df(x^{(0)})x^{(1)} + g(x^{(0)}).
\end{equation}
The general solution to Eq.(\ref{eq:1st-eq}) is composed of the
homogeneous solution and a special solution.
The homogeneous solution has extra $n$ integration constants,
which are written as the $n$-dimensional column vector $a$,
and is denoted by $\phi^{(1)}(t;A,a)$ accordingly. 
The homogeneous solution can be rewritten as
\begin{equation}
  \label{eq:phi1-phi0}
  \phi^{(1)}(t;A,a) = \dfracp{\phi^{(0)}}{A}(t;A) a,
\end{equation}
where $\partial \phi^{(0)}/\partial A$ is the Jacobi matrix.
To show this equation,
we note that the function $\phi^{(0)}(t;A+\ep a)$ also satisfies
the first order equation: 
$(\partial\phi^{(0)}/\partial t)(t;A+\ep a)
=f(\phi^{(0)}(t;A+\ep a))$.
Comparing the left- and the right-hand-sides
in terms of the first order of the Taylor series,
we find that $(\partial\phi^{(0)}/\partial A) a$
is the general solution to the homogeneous part of
Eq.(\ref{eq:1st-eq}).
The special solution is the sum of an arbitrary homogeneous solution,
$\phi^{(1)}(t;A,a')$, and the inhomogeneous solution, 
which has no extra constants and is denoted by $\psi(t;A)$.
Since the homogeneous solution is not affected by the perturbation
$g(x)$,
secular terms are included by $\psi(t;A)$ if they exist.
We therefore decompose it into two ports:
$\psi(t;A)=\psi^{(s)}(t;A)+\psi^{(ns)}(t;A)$,
where $\psi^{(s)}(t;A)$ and $\psi^{(ns)}(t;A)$
consist of secular and nonsecular terms, respectively.
In the end, the naive but general solution up to the first order is
formally written as 
\begin{equation}
  \label{eq:naive-sol}
  x^{NS}(t)=\phi^{(0)}(t;A) + \ep [ \phi^{(1)}(t;A,a) +
  \psi^{(s)}(t;A) + \psi^{(ns)}(t;A)],
\end{equation}
where we have replaced $a+a'$ with $a$.

The first and crucial step for constructing 
the unique renormalized solution
is to determine a local solution
from an initial condition, $x^{NS}(t_{0})=x_{0}$,
where the local solution must be valid around $t_{0}$.
Let $A_{0}$ and $a_{0}$ be the values of $A$ and $a$ 
determined by the initial condition, respectively.
The number of elements of $A_{0}$ and $a_{0}$ is $2n$,
while the initial condition can determines values of $n$ elements
among them. 
We thus introduce an assumption that
the extra integration constants, $a$, cancel out
all the secular terms of the naive solution (\ref{eq:naive-sol})
at $t=t_{0}$:
\begin{equation}
  \label{eq:condition-a-t0}
  \phi^{(1)}(t_{0};A_{0},a_{0}) 
  + \psi^{(s)}(t_{0};A_{0}) = 0.
\end{equation}
This condition uniquely determines the value $a_{0}$
from $A_{0}$ and $t_{0}$ unless accidental degeneracy occurs.
The local solution being valid around $t_{0}$, then, becomes
\begin{equation}
  \label{eq:local-sol-t0}
  \begin{split}
    x^{NS}(t;t_{0}) 
    & = \phi^{(0)}(t;A_{0})
    + \ep [ \phi^{(1)}(t;A_{0},a_{0}) \\
    & + \psi^{(s)}(t;A_{0}) + \psi^{(s)}(t;A_{0})],
  \end{split}
\end{equation}
and the value $A_{0}$ is uniquely determined from the initial
condition $x^{NS}(t_{0};t_{0})=x_{0}$ with
Eq.(\ref{eq:condition-a-t0}).

The second step is to connect the local solution $x^{NS}(t;t_{0})$ with
another local solution being valid around $\mu$.
The time $\mu$ may be arbitrarily chosen from a suitable
neighborhood of $t_{0}$ where the local solution $x^{NS}(t;t_{0})$ is
valid, 
and hence we regard the integration constant $A$ as a function of
$\mu$, and denote it by $A(\mu)$.
Similarly, the extra integration constant $a$ is regarded as a
function of $A(\mu)$ and $\mu$,
and is denoted by $a(A(\mu),\mu)$.
We remark that $A(t_{0})=A_{0}$ and $a(A(t_{0}),t_{0})=a_{0}$.
Then the local solution around $\mu$ is similar to
Eq.(\ref{eq:local-sol-t0}): 
\begin{displaymath}
  \begin{split}
    x^{NS}(t;\mu) 
    & = \phi^{(0)}(t;A(\mu))
    + \ep [ \phi^{(1)}(t;A(\mu),a(A(\mu),\mu)) \\
    & + \psi^{(s)}(t;A(\mu)) + \psi^{(ns)}(t;A(\mu))],
  \end{split}
\end{displaymath}
and $a(A(\mu),\mu)$ must be satisfied the similar
condition to Eq.(\ref{eq:condition-a-t0}):
\begin{equation}
  \label{eq:condition-a}
  \phi^{(1)}(\mu;A(\mu),a(A(\mu),\mu) )
  + \psi^{(s)}(\mu;A(\mu)) = 0.
\end{equation}
Using Eq.(\ref{eq:phi1-phi0}),
the values of the extra integration constants are formally determined
as
\begin{equation}
  \label{eq:form-a}
  a(A(\mu),\mu) = 
  - \left(
    \dfracp{\phi^{(0)}}{A}(\mu;A(\mu))
  \right)^{-1} \psi^{(s)}(\mu;A(\mu)).
\end{equation}
The connection condition, $x^{NS}(t;\mu)=x^{NS}(t;t_{0})$,
is satisfied up to the first order of $\ep$
by putting $A(\mu)=A(t_{0})+\delta A$ and deciding the $\delta A$
as 
\begin{displaymath}
  \delta A = - \ep [ a(A(t_{0}),\mu) - a(A(t_{0}),t_{0}) ].
\end{displaymath}

Remembering $\delta A=A(\mu)-A(t_{0})$
and taking the limit $\mu\to t_{0}$,
we obtain the relation between differential coefficients at $t_{0}$:
$\rmd A/\rmd \mu (t_{0}) = - \ep \partial a/\partial \mu
(A(t_{0}),t_{0})$.
Since the initial time $t_{0}$ may be chosen arbitrarily,
we regard the relation as the differential equation of the form
\begin{displaymath}
  \dfracd{A}{\mu}(\mu) = - \ep \dfracp{a}{\mu}(A(\mu),\mu).
\end{displaymath}
This is the renormalization group equation,
and must be solved with the initial condition $A(t_{0})=A_{0}$.
We remark that the renormalization group equation is determined
by the extra integration constants through the dependence on $A$ and
$\mu$,
which is assumed as Eq.(\ref{eq:form-a}).

Finally, the renormalized solution, $x^{RG}(t)$, is obtained
by evaluating the naive solution $x^{NS}(t;t_{0})$ at $t_{0}=t$
and $A(t_{0})=A(t)$,
because we are interested in the solution being valid around
the time $t$.
The renormalized solution is, therefore, expressed as
\begin{equation}
  \label{eq:RGS}
  x^{RG}(t)=x^{NS}(t;t)
  = \phi^{(0)}(t;A(t)) + \ep \psi^{(ns)}(t;A(t)),
\end{equation}
by using the condition (\ref{eq:condition-a}).

Let us apply this procedure to the following three examples.
We use the same symbols introduced above,
and denote elements of $x$, $A$ and $a$ by
$x_{j}$, $A_{j}$ and $a_{j}$, $j=1,2$, respectively.

\vspace*{1em}
\noindent
{\bf Example 1} - Harmonic oscillator \\
The first example is a harmonic oscillator with linear perturbation:
\begin{displaymath}
  \dfracd{}{t}
  \begin{pmatrix}
    x_{1} \\ x_{2}
  \end{pmatrix}
  =
  \begin{pmatrix}
    0 & 1 \\ -1 & 0
  \end{pmatrix}
  \begin{pmatrix}
    x_{1} \\ x_{2}
  \end{pmatrix}
  + \ep
  \begin{pmatrix}
    0 \\ -x_{1}
  \end{pmatrix} .
\end{displaymath}
First, we diagonalize the matrix found in the right-hand-side
of this equation with the diagonalizing matrix $R$ and
the coordinate transformation which are
\begin{displaymath}
  R =
  \begin{pmatrix}
    1 & i \\ i & 1
  \end{pmatrix},
  \quad
  \begin{pmatrix}
    x_{1} \\ x_{2}
  \end{pmatrix}
  = R
  \begin{pmatrix}
    y_{1} \\ y_{2}
  \end{pmatrix} ,
\end{displaymath}
respectively.
The original equation is transformed as
\begin{equation}
  \label{eq:example-1}
  \dfracd{}{t}
  \begin{pmatrix}
    y_{1} \\ y_{2}
  \end{pmatrix}
  =
  \begin{pmatrix}
    i & 0 \\ 0 & -i 
  \end{pmatrix}
  \begin{pmatrix}
    y_{1} \\ y_{2}
  \end{pmatrix}
  + \dfrac{\ep}{2}
  \begin{pmatrix}
    i y_{1} - y_{2} \\ - y_{1} - i y_{2}
  \end{pmatrix} .
\end{equation}
The naive solutions up to the first order are calculated as
\begin{equation}
  \label{eq:ns-example1}
  \begin{split}
    y_{1}^{NS}(t) & = A_{1} e^{it} + \ep \left[
      a_{1} e^{it} + \dfrac{i}{2} A_{1} t e^{it} 
      + \dfrac{1}{4i} A_{2} e^{-it}  \right], \\
    y_{2}^{NS}(t) & = A_{2} e^{-it} + \ep \left[
      a_{2} e^{-it} - \dfrac{i}{2} A_{2} t e^{-it} 
      - \dfrac{1}{4i} A_{1} e^{it}  \right], \\
  \end{split}
\end{equation}
and $\phi^{(1)}$ and $\psi^{(s)}$ are determined as
\begin{displaymath}
  \phi^{(1)}(t;A,a) =
  \begin{pmatrix}
    a_{1} e^{it} \\ a_{2} e^{-it}
  \end{pmatrix},
  \quad
  \psi^{(s)}(t;A) = \dfrac{i}{2} 
  \begin{pmatrix}
    A_{1} t e^{it} \\ - A_{2} t e^{-it}
  \end{pmatrix},
\end{displaymath}
respectively. From the condition (\ref{eq:condition-a}),
the extra integration constants, $a_{1}$ and $a_{2}$, 
are determined as
\begin{equation}
  \label{eq:a-example1}
  a_{1}(A(\mu),\mu) = - i A_{1}(\mu)\mu / 2 ,
  \quad
  a_{2}(A(\mu),\mu) = i A_{2}(\mu)\mu / 2,
\end{equation}
and the renormalization group equations are
\begin{displaymath}
  \dfracd{A_{1}}{\mu} = \ep \dfrac{i}{2} A_{1},
  \quad
  \dfracd{A_{2}}{\mu} = - \ep \dfrac{i}{2} A_{2}.
\end{displaymath}
Solving these renormalization group equations
with the initial condition
$(A_{1}(t_{0}),A_{2}(t_{0}))=(A_{1,0},A_{2,0})$ , 
which is determined by Eqs.(\ref{eq:ns-example1}),(\ref{eq:a-example1})
and the initial condition
$(y_{1}(t_{0}),y_{2}(t_{0})) = (y_{1,0},y_{2,0})$,
the renormalized solutions are uniquely obtained as
\begin{displaymath}
  \begin{split}
    y_{1}^{RG}(t) & = 
    A_{1,0}~ e^{i\ep (t-t_{0})/2} e^{it}
    + \dfrac{\ep}{4i} A_{2,0}~ e^{-i\ep (t-t_{0})/2} e^{-it},\\
    y_{2}^{RG}(t) & =
    A_{2,0}~ e^{-i\ep (t-t_{0})/2} e^{-it}
    - \dfrac{\ep}{4i} A_{1,0}~ e^{i\ep (t-t_{0})/2} e^{it}.
  \end{split}
\end{displaymath}
Note that the renormalized solutions satisfy the original 
equation (\ref{eq:example-1}) up to the first order,
and the frequency of the renormalized solution, $1+\ep/2$,
approximates the frequency of the exact solution, $\sqrt{1+\ep}$,
up to the first order of the Taylor series accordingly.

\vspace*{1em}
\noindent
{\bf Example 2} - Jordan cell\\
The second example is an equation having a Jordan cell in the
unperturbed part:
\begin{equation}
  \label{eq:example-2}
  \dfracd{}{t}
  \begin{pmatrix}
    x_{1} \\ x_{2}
  \end{pmatrix}
  =
  \begin{pmatrix}
    0 & 1 \\ 0 & 0
  \end{pmatrix}
  \begin{pmatrix}
    x_{1} \\ x_{2}
  \end{pmatrix}
  + \ep
  \begin{pmatrix}
    0 \\ - x_{1}
  \end{pmatrix} .
\end{equation}
The naive solutions up to the first order are calculated as
\begin{displaymath}
  \begin{split}
    x_{1}^{NS}(t) & = 
    A_{1} + A_{2}t + \ep \left[ a_{1} + a_{2} t
      - \dfrac{1}{2} A_{1} t^{2} - \dfrac{1}{3!} A_{2} t^{3} \right],
    \\  
    x_{2}^{NS}(t) & =
    A_{2} + \ep \left[ a_{2} - A_{1} t - \dfrac{1}{2} A_{2} t^{2}
    \right].
  \end{split}
\end{displaymath}
The solutions of the zeroth order do not oscillate,
and no resonance occurs accordingly.
Consequently, it is not obvious which are secular terms,
but we regard the last two terms in each of
$x_{1}^{NS}(t)$ and $x_{2}^{NS}(t)$ as the secular terms
since their absolute values exceed absolute values of zeroth order
when $\ep t \gtrsim 1$.
To vanish all the secular terms simultaneously at $t=\mu$,
the extra integration constants, $a_{1}$ and $a_{2}$, must be
determined as 
\begin{displaymath}
  \begin{split}
    a_{1}(A(\mu),\mu) & = -A_{1}(\mu) \mu^{2}/2
    - A_{2}(\mu) \mu^{3}/3, \\
    a_{2}(A(\mu),\mu) & = A_{1}(\mu)+A_{2}(\mu)\mu^{2}/2.
  \end{split}
\end{displaymath}
Then the renormalization group equations are
\begin{displaymath}
  \dfracd{A_{1}}{\mu} = \ep (\mu A_{1} + \mu^{2} A_{2} ),
  \quad
  \dfracd{A_{2}}{\mu} = -\ep ( A_{1} + \mu A_{2} ).
\end{displaymath}
We emphasize that the renormalized solutions,
$x_{1}^{RG}(t)=A_{1}(t)+A_{2}(t)t$ and $x_{2}^{RG}(t)=A_{2}(t)$,
satisfy the original equation (\ref{eq:example-2})
with the aid of the renormalization group equations,
and hence they are exact solutions.

We remark that an equation which has a Jordan cell in its unperturbed
part is also discussed in Ref.\cite{ei-00},
but one of the renormalized integral constants is selected
from a homogeneous solution of the first order.

\vspace*{1em}
\noindent
{\bf Example 3} - Nonlinear equation \\
The third example is an equation whose unperturbed part is nonlinear:
\begin{equation}
  \label{eq:example-3}
  \dfracd{}{t}
  \begin{pmatrix}
    x_{1} \\ x_{2}
  \end{pmatrix}
  =
  \begin{pmatrix}
    x_{2} \\ - 3 x_{2}^{2}/2
  \end{pmatrix}
  - \dfrac{\ep}{2}
  \begin{pmatrix}
    0 \\ - \exp (-2x_{1})
  \end{pmatrix} . 
\end{equation}
This is Einstein's equation for a
Friedman-Robertson-Walker universe with dust,
and $x_{1}(t)$ is the logarithm of the scale factor of the universe 
and $\ep$ is the sign of the spatial curvature.
The naive solutions up to the first order are
\begin{displaymath}
  \begin{split}
    x_{1}^{NS}(t) & = \ln(t+A_{1})^{2/3} + A_{2} \\
    & + \ep \left[ a_{1} (t+A_{1})^{-1} + a_{2}
      - \dfrac{9}{20} (t+A_{1})^{2/3} \exp(-2A_{2}) \right], \\
    x_{2}^{NS}(t) & = \dfrac{2}{3} (t+A_{1})^{-1} \\
    & + \ep \left[ - a_{1} (t+A_{1})^{-2}
      - \dfrac{3}{10} (t+A_{1})^{-1/3} \exp(-2A_{2}) \right].
  \end{split}
\end{displaymath}
The solutions of the zeroth order do not oscillate again,
and hence secular terms are not determined automatically.
If we decide the secular term as
\begin{displaymath}
  \psi^{(s)}(t;A) =
  \begin{pmatrix}
    - 9 (t+A_{1})^{2/3} \exp(-2A_{2}) / 20 \\ 0
  \end{pmatrix},
\end{displaymath}
then the renormalized solution coincides with one obtained in
Ref.\cite{nambu-99}
with changing the independent variable as $\tau=(\mu+A_{1})^{2/3}$.
The renormalized solution $x_{1}^{RG}(t)$ improves the naive solution
$x_{1}^{NS}(t)$ and reproduces expanding and contracting feature of
the exact solution.
On the other hand, if we decide the secular terms as
\begin{displaymath}
  \psi^{(s)}(t;A) =
  \begin{pmatrix}
    - 9 (t+A_{1})^{2/3} \exp(-2A_{2}) / 20 \\
    - 3 (t+A_{1})^{-1/3} \exp(-2A_{2}) / 10
  \end{pmatrix},
\end{displaymath}
then the renormalization group equations are
\begin{displaymath}
  \begin{split}
    \dfracd{A_{1}}{\mu} & = \ep \dfrac{3}{4} e^{-2A_{2}}
  (\mu+A_{1})^{2/3}, \\
  \dfracd{A_{2}}{\mu} & = - \ep \dfrac{1}{2} e^{-2A_{2}}
  (\mu+A_{1})^{-1/3},
  \end{split}
\end{displaymath}
and the renormalized solutions
satisfy the original equation (\ref{eq:example-3}) exactly.

\vspace*{1em}
In Examples 2 and 3,
renormalized solutions become exact 
when nonsecular terms do not appear in the first order.
In the following Theorem and Corollary,
we show that the feature holds in general.
As a preparation of the Theorem,
we define the functions $g^{(\gamma)},\gamma=s,ns,$ as
\begin{displaymath}
  g^{(\gamma)}(\phi^{(0)}(t;A))
  = \dfracp{\psi^{(\gamma)}}{t}(t;A) - Df(\phi^{(0)}(t;A))
  \psi^{(\gamma)}(t;A),
\end{displaymath}
and the symbol $\lessless f(x^{RG}) \moremore_{2}$ as
the higher orders of the Taylor series of $f(x^{RG})$, i.e.
\begin{displaymath}
  \lessless f(x^{RG}) \moremore_{2}
  = \sum_{k=2}^{\infty} \dfrac{[\ep \psi^{(ns)}(t;A(t))]^{k}}{k!}
  f^{(k)}(\phi^{(0)}(t;A(t))) ,
\end{displaymath}
where $f^{(k)}$ represents the $k$-th order derivative of $f$.

\vspace*{1em}
\noindent
{\bf Theorem:}
Let $x^{RG}(t)$ be a renormalized solution to
the ordinally differential equation
(\ref{eq:original-eq}) up to the first order.
Then the following relation holds:
\begin{equation}
  \label{eq:theorem}
  \begin{split}
    & \dfracd{x^{RG}}{t}(t) 
    - \left[ f(x^{RG}(t)) + \ep g(x^{RG}(t)) \right] \\
    & = \ep^{2} \left[ 
      \left( \dfracp{\psi^{(ns)}}{A}(t;A(t)) \right)
      \left( \dfracp{\phi^{(0)}}{A}(t;A(t)) \right)^{-1} \right. \\
    &  \left. \ g^{(s)}(\phi^{(0)}(t;A(t))) 
      - Dg(\phi^{(0)}(t;A(t))) \psi^{(ns)}(t;A(t)) \right] \\
    & \hspace*{1em} - \lessless f(x^{RG}) \moremore_{2}
    - \ep \lessless g(x^{RG}) \moremore_{2}. 
  \end{split}
\end{equation}

\vspace*{1em}
\noindent
{\bf Proof:}
In this proof, we omit arguments of functions
which are evaluated at $\mu=t$ and $A=A(t)$.
Note that, from the definitions, the relations hold:
$\partial \phi^{(0)}/\partial t = f(\phi^{(0)}),\ 
\partial \psi^{(ns)}/\partial t = Df(\phi^{(0)}) \psi^{(ns)} +
g^{(ns)}(\phi^{(0)})$ and so on.

Derivating the renormalized solution (\ref{eq:RGS}) with respect to
$t$, we obtain the following equation
\begin{displaymath}
  \begin{split}
    \dfracd{x^{RG}}{t}
    & = f( \phi^{(0)} ) 
    - \ep \dfracp{\phi^{(0)}}{A} \dfracp{a}{\mu} \\
    & + \ep \left[ Df( \phi^{(0)} ) \psi^{(ns)}
      + g^{(ns)}( \phi^{(0)} ) \right]
    - \ep^{2} \dfracp{\psi^{(ns)}}{A} \dfracp{a}{\mu}.
  \end{split}
\end{displaymath}
Next, partially derivating Eq.(\ref{eq:condition-a}) with respect to
$\mu$, we find the equation of the form
\begin{displaymath}
  \dfracp{\phi^{(0)}}{A} \dfracp{a}{\mu} + g^{(s)}(\phi^{(0)}) = 0,
\end{displaymath}
where we used Eqs.(\ref{eq:phi1-phi0}) and (\ref{eq:condition-a}).
From the two equations mentioned above,
we obtain the following relation 
\begin{displaymath}
  \begin{split}
    \dfracd{x^{RG}}{t}
    & = f( \phi^{(0)} )  + \ep g( \phi^{(0)} )
    + \ep Df( \phi^{(0)} ) \psi^{(ns)} \\
    & + \ep^{2}   \left( \dfracp{\psi^{(ns)}}{A} \right)
    \left( \dfracp{\phi^{(0)}}{A} \right)^{-1}
    g^{(s)}(\phi^{(0)} ) .
  \end{split}
\end{displaymath}
On the other hand, from Eq.(\ref{eq:RGS}),
the Taylor series of $f(x^{RG}(t))+\ep g(x^{RG}(t))$ becomes
\begin{displaymath}
  \begin{split}
    & f(x^{RG})+\ep g(x^{RG})
    = f(\phi^{(0)}) + \ep g(\phi^{(0)}) 
    +\ep Df(\phi^{(0)}) \psi^{(ns)} \\
    & + \ep^{2} Dg(\phi^{(0)}) \psi^{(ns)}
    + \lessless f(x^{RG}) \moremore_{2}
    + \ep \lessless g(x^{RG}) \moremore_{2} .
  \end{split}
\end{displaymath}
Consequently, we obtain the relation stated in the theorem.

\vspace*{1em}
\noindent
{\bf Corollary:}
The renormalized solution up to the first order is the exact solution
to the original equation
if the naive solution of the first order does not include nonsecular
terms. 

\vspace*{1em}
\noindent
{\bf Proof:}
From the assumption, $\psi^{(ns)}=0$, and 
$\lessless f(x^{RG}(t) \moremore_{2}=
\lessless g(x^{RG}(t)) \moremore_{2}=0$ accordingly.
Consequently, the right-hand-side of Eq.(\ref{eq:theorem}) vanishes,
and the renormalized solution $x^{RG}(t)$ exactly satisfies the
original equation.

As summary, we propose a procedure for uniquely constructing 
a renormalized solution from a given initial condition.
The basic idea of the procedure is to determine extra integral
constants as they cancel out all the secular terms simultaneously
at the initial time.
As a result of the procedure,
the theorem is stated that a renormalized solution
is the exact solution to an original equation
if the naive solution does not include nonsecular terms
in the first order of a small parameter.
Note that renormalization group equations are not always autonomous
even if original equations are.

The renormalization group method is also available to obtain reduced
equations as the renormalization group equations
\cite{matsuba-97a,matsuba-97b,goto-01b}.
The procedure proposed in this letter clarifies the relation
between the renormalization group equation with the way how we decide
secular terms,
and hence it seems useful not only obtaining approximate solutions
but also reducing original equations.

The author would like to thank K.Nozaki, K.~Imai, Y.~Nambu 
and A.~Taruya for fruitful discussions and comments.
This work is supported by the Grand-In-Aid for Scientific Research of
the Ministry of Education, Culture, Sports, Science and Technology
of Japan (12750060).

\end{document}